\documentclass[aps,pre,reprint]{revtex4-2}

\usepackage[utf8]{inputenc}

\usepackage{amsmath}
\usepackage{amssymb}
\usepackage{amsfonts}
\usepackage{graphicx} 
\usepackage{hyperref} 
\usepackage{xcolor}
\usepackage[normalem]{ulem}

\begin{document}

\title{Memory-induced long-range order in dynamical systems}

\author{Chesson Sipling}
\email{email: csipling@ucsd.edu}
\affiliation{Department of Physics, University of California San Diego, La Jolla, CA 92093}

\author{Yuan-Hang Zhang}
\email{email: yuz092@ucsd.edu}
\affiliation{Department of Physics, University of California San Diego, La Jolla, CA 92093}

\author{Massimiliano Di Ventra}
\email{email: diventra@physics.ucsd.edu}
\affiliation{Department of Physics, University of California San Diego, La Jolla, CA 92093}

\begin{abstract}
Time non-locality, or memory, is a non-equilibrium property shared by all physical systems. Here, we show that memory is sufficient to induce a {\it phase} of spatial long-range order (LRO) even if the system’s primary dynamical variables are coupled locally. This occurs when the memory degrees of freedom have slower dynamics than the primary degrees of freedom. In addition, such an LRO phase is {\it non-perturbative}, and can be understood through the lens of a correlated percolation transition of the fast degrees of freedom mediated by memory. When the two degrees of freedom have comparable time scales, the length of the effective long-range interaction shortens. We exemplify this behavior with a model of locally coupled spins and a single dynamic memory variable, but our analysis is sufficiently general to suggest that memory could induce a phase of LRO in a much wider variety of physical systems. 
\end{abstract}


\maketitle

\section{Introduction}
Memory, or {\it time non-locality}, is a general property shared by all physical systems, both classical and quantum~\cite{kubo1957statistical}. It means that the response of a physical system to a perturbation at a given time depends on the full history of its past dynamics. Physically, memory originates from the dynamics of some degrees of freedom, such as the spin polarization in a magnetic material, the motion of atomic vacancies in a solid, etc.~\cite{11_memory_materials}. Although it is a ubiquitous property, it is less obvious how it affects {\it spatial} interactions between the components of a system. A simple example from Nature may provide some illuminating clues to this question.

Imagine a maze that separates a nest of ants from a feeding site. It is well known that as the ants traverse the maze, they will excrete pheromones~\cite{wilson1963social} that persist for some time. These pheromones act as memory degrees of freedom (DOFs), providing data that inform how future ants navigate the maze's complex landscape~\cite{10.1242/jeb.076570}. As more ants traverse the maze in search of food, the optimal path is reinforced by the pheromone trail. Not only does this allow ants to communicate non-locally in time, but also in {\it space}, as ants that are otherwise very far from one another can still influence each other's dynamics due to the persistence of the pheromones.

In this example, there are two distinct DOFs (ants and pheromones) that evolve over different time scales. The pheromone DOFs decay much more slowly than a typical ant's maze-traversal time, so the memory can persist and induce strong correlations between spatially distant ants which would otherwise be weakly correlated or completely uncorrelated. In other words, an {\it effective} spatial long-range correlation between the ants is established due to the influence of the pheromones (memory).

While many studies have commented on the relationship between non-equilibrium dynamics and spatial long-range order (LRO)~\cite{hinrichsen2000non, dewar2003information, tauber2014critical, PEI2022126727, caravelli, PhysRevLett.124.010603}, to the best of our knowledge none have detailed an explicit connection between time non-locality and spatial LRO~\footnote{However, such a connection is practically exploited in the MemComputing paradigm of computation~\cite{di2022memcomputing}.}. In this work, we hope to remedy this by revealing how time non-locality can induce LRO in general physical systems. Throughout this article we use LRO to denote correlations that decay at most as a power-law with distance or time. When correlations do not decay, we describe the state as ``rigid''.

First, we will perform some general analysis to demonstrate how the coupling between two DOFs with sufficiently distinct timescales can induce long-range correlations in the faster DOFs, even when couplings are explicitly local. In particular, such LRO exists over a wide range of parameters and hence constitutes a {\it phase} that is not necessarily associated with criticality. As we will show, this phase is {\it non-perturbative}, meaning that the fast and slow DOFs become strongly coupled soon after the dynamics are initiated. To elucidate this analysis, we will concentrate on a particular numerical model: a lattice of (continuously relaxed) spins in the presence of a memory field. We will simulate this lattice system for many different values of the memory timescale, extract ensemble-averaged avalanche distributions, and analyze the system through the lens of correlated percolation. Finally, we will summarize our results and the implications of our work for general dynamical systems.

\section{General Analysis}
Consider the following lattice system in which two DOFs (a position-dependent ``primary'' DOF, $s_i$, and an interaction-dependent ``memory'' DOF, $x_{i j}$) are coupled locally:

\begin{equation}
\dot{\bf x}={\bf F}({\bf x})\equiv\left\{
\begin{aligned}
    \dot{s_i} &= g \sum_{<i j>} x_{i j} s_i\,, \\
    \dot{x}_{i j} & = \gamma s_i s_j\,, \label{eq:simplem}
\end{aligned}
\right
.
\end{equation} 

where ${\bf x} =\{ s_i, {x}_{i j}\}$, ${\bf F}$ is the flow field, and the coupling of strength $g$ between the $s_i$'s is nearest-neighbor.

We will restrict $s_i$ and $x_{i j}$ so that $s_i \in [-1,1]$ and $x_{i j} \in [0,1]$, forming a compact phase space. In other words, if the flow field in Eq.~(\ref{eq:simplem}) were to push a DOF outside of its bounds, the system instead truncates it onto its bounds. We can prescribe $1/g$ and $1/\gamma$ as characteristic $s_i$ and $x_{i j}$ timescales, respectively. For the $x_{i j}$ to act as {\it memory} DOFs for the $s_i$'s, their dynamics must be slower than the $s_i$'s dynamics (i.e., $g > \gamma$). Thus, the memory DOFs will not vary much over timescales less than $1/\gamma$, and similarly, the primary DOFs will vary little over timescales less than $1/g$.

This allows us to perform an iterative integration scheme to find an approximate form for $s_i(t = T)$. A similar technique has been used for complex, spatially coupled systems in other various fields from transport theory \cite{10.1063/1.1724117} to neural dynamics \cite{wilson_cowan}. We found this approach more successful than multiple-scale analysis \cite{17039}, which we found both analytically untenable and unnecessary for our system (as we are not so interested in finding closed-form solutions for $s_i(t)$ and $x_{ij}(t)$).

We must emphasize that the following analysis is {\it not} exact and will only be used to make predictions about the system's correlation structure. In particular, the boundedness of $s_i(t)$ and $x_{ij}(t)$ is not fully accounted for in Eqns. (\ref{eq:si1}), (\ref{eq:xij}), and (\ref{eq:si2}) (both in their integrands and their RHS's, which implicitly, must also be truncated). Nonetheless, we do not suspect these caveats significantly affect our results.

We will work in the limit $g \gg \gamma$. If we discretize time into units of length $1/\gamma$, we can evaluate the primary DOFs at the first time step $T = 1/\gamma$ as

\begin{equation}
\begin{split}
    s_i(1/\gamma) &\approx s_i(0) + \int_0^{1/\gamma} dt \, g \sum_{<i j'>} x_{i j'}(t) s_i(t) \\
    & \approx s_i(0) + \frac{g}{\gamma} \sum_{<i j'>} x_{i j'}(0) \overline{s_{i, 0}}\,, \label{eq:si1}
\end{split}
\end{equation} 

where we have defined

\begin{equation}
    \overline{s_{i, l}} \equiv \frac{1}{n} \sum_{a = 0}^{n - 1} s_i(l/\gamma + a/g)\,,
\end{equation} 

and have assumed that the memory DOFs vary little during this interval.

$\overline{s_{i, l}}$ is the average value of $s_i$ between $t = l/\gamma$ and $t = (l+1)/\gamma$; $n$ is the ratio of the slow, memory DOF timescale to the fast, primary DOF timescale (in this case, $n \equiv \lfloor g/\gamma \rfloor$, where $\lfloor \cdot \rfloor$ is the floor function). Similarly, at time $T = 1/\gamma$, the memory DOFs take the form

\begin{equation}
\begin{split}
    x_{i j}(1/\gamma) &\approx x_{i j}(0) + \int_0^{1/\gamma} dt \, \gamma s_i(t) s_j(t) \\
    & \approx x_{i j}(0) + \overline{s_{i, 0}} \overline{s_{j, 0}} \\
    & \approx x_{i j}(0) + \bigg(s_i(0) + \frac{g}{\gamma} \sum_{<i j'>} x_{i j'}(0) \overline{s_{i, 0}} \bigg) \\
    & \quad \quad \times \bigg(s_j(0) + \frac{g}{\gamma} \sum_{<j k'>} x_{j k'}(0) \overline{s_{j, 0}} \bigg)\,. \label{eq:xij}
\end{split}
\end{equation} 

In the third line of Eq.~(\ref{eq:xij}), we have inserted $s_i(1/\gamma)$ from Eq.~(\ref{eq:si1}) as an approximation of $\overline{s_{i, 0}}$. Note the emergence of the index $k'$, representing the {\it next}-nearest-neighbors of $s_i$. We can then evaluate the primary DOFs at the next time step $T = 2/\gamma$:

\begin{equation}
\begin{split}
    s_i(2/\gamma) &\approx s_i(1/\gamma) + \int_{1/\gamma}^{2/\gamma} dt \, g \sum_{<i j'>} x_{i j'}(t) s_i(t) \\
    & \approx s_i(1/\gamma) + \frac{g}{\gamma} \sum_{<i j'>} x_{i j'}(1/\gamma) \overline{s_{i, 1}} \\
    & \approx \bigg(\frac{g}{\gamma}\bigg)^2 s_i(0) \sum_{<i j'>} \sum_{<j' k'>} x_{j' k'}(0) \overline{s_{j', 0}} \, \overline{s_{i, 1}}\,+\dots. \label{eq:si2}
\end{split}
\end{equation}\ 

In Eq.~(\ref{eq:si2}), we only show the term of order $(\mathrm{g}/\gamma)^2$ that explicitly couples the DOFs $s_i$ to their nearest-neighbors $s_{j'}$. Continuing this approach will yield higher order terms that couple $s_i$'s further away in the lattice (e.g., we expect a term $\propto (g/\gamma)^3 \sum_{<i j'>} \sum_{<j' k'>} \sum_{<k' l'>} x_{k' l'}(0) \overline{s_{k', 0}} \, \overline{s_{j', 1}} \, \overline{s_{i, 2}}$ after one more iteration).


Crucially, we recognize this effect as {\it non-perturbative} since the parameter $g/\gamma>> 1$ by assumption, and each additional order in coupling produces terms with additional factors of $g/\gamma$. This means that the correlations {\it cannot} decay exponentially; the long-range terms $\propto (g/\gamma)^d$, where $d$ is the number of lattice sites between coupled DOFs, will contribute {\it strongly} to the system's dynamics. This situation is distinct from other systems in which implicit, beyond nearest-neighbor couplings whose contributions decay exponentially emerge (which is not something we would characterize as ``long-range''). In the supplementary material (SM), we show how the memory DOFs also correlate at long distances, acting as an effective long-range potential for the spins.

All this shows that, given a long enough time, we expect LRO in the primary (fast) DOFs to emerge due to the presence of strong, high-order, beyond-nearest-neighbor couplings. In other words, {\it time non-locality in coupled DOFs can induce long-range correlations}. In our analysis, the memory DOFs induce an {\it effective} LRO in the primary DOFs, in much the same way that the slow decay of ant pheromones can induce strong non-local coupling amongst a population of ants.

Furthermore, this effect is very robust against perturbations in either the primary or memory DOFs. Since each DOF is coupled explicitly to multiple adjacent neighbors, there are multiple ``paths'' through which the memory can couple spatially distant primary DOFs. So, a local perturbation (e.g., the breaking of some links between the $s_i$'s in the lattice, so long as the connectivity of the lattice is unchanged) should not destroy this effective LRO. This is again analogous to the ants in the maze: with the help of the time non-local pheromones, the ants would be able to circumvent new perturbative obstacles/changes in the maze and still reach the food.

This analysis holds also in the continuous case ($s_i \to s({\bf r})$ and $x_{ij} \to x({\bf r}_1, {\bf r}_2)$), so long as the memory timescale is still relatively long and there is some local (diffusive) coupling present between the nearby primary DOFs.

\section{Spin glass-inspired model}
To study this phenomenon concretely, we examine a spin-glass-like model coupled to a memory field on a 2D square lattice with periodic boundary conditions. We note that, although this model has additional structures compared to the one in the previous section, the form of spin-memory coupling is identical and the previous analysis still holds. Therefore, we still expect LRO to emerge amongst the spins due to their coupling to the slow memory DOFs. We choose the following spin glass Hamiltonian:

\begin{figure}
    \includegraphics[width=\columnwidth]{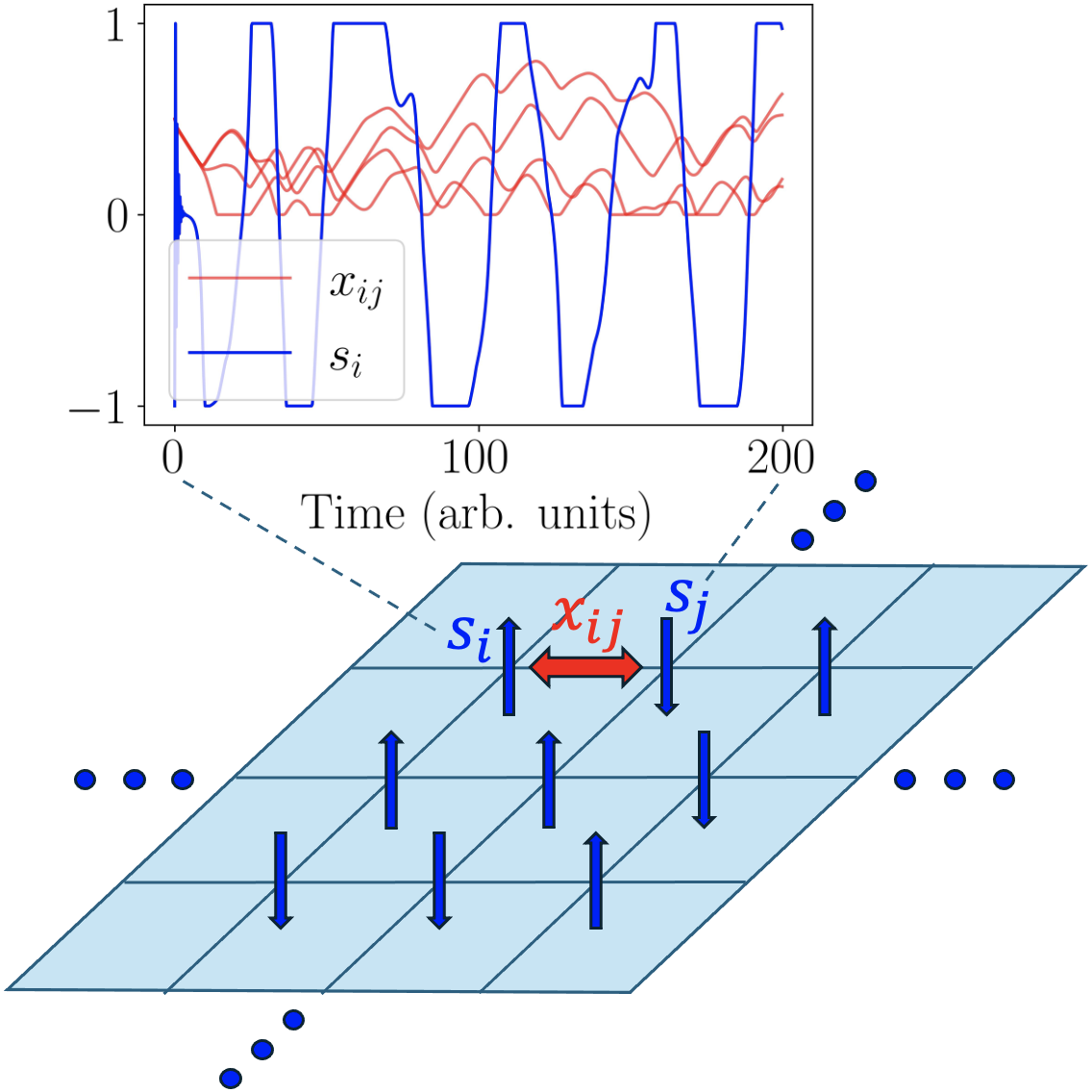}
    \caption{Cartoon of a continuous spin system with memory on a 2D lattice. Nearest-neighbor spins $s_i$ and $s_j$ are coupled by the memory DOF $x_{i j}$. Dynamics from Eq.~(\ref{eq:memory_dynamics}) are simulated in a square lattice of size $N = 16^2$ with $\gamma = 0.1$. For a randomly chosen site $i$, $s_i(t)$ (blue) and its four $x_{i j}(t)$ (red) are plotted. \label{fig:spin_glass_viz}} 
\end{figure}

\begin{equation}
    H = -\sum_{<i j>} J_{i j} s_i s_j\,.
\end{equation} 

Here, the $J_{i j} = \pm 1$ nearest-neighbor interactions are symmetric, but otherwise randomly distributed throughout the lattice. If we continuously relax these spins, such that $s_i \in [-1, 1]$ rather than $s_i \in \{-1, 1\}$, we obtain the spin dynamics:

\begin{equation}
    \dot{s}_i = -\nabla_ {{s}_i} H=\sum_{<i j>} J_{i j} s_j\,. \label{eq:glass_dynamics}
\end{equation} 

As in Eq.~(\ref{eq:simplem}), we now couple these spins to a memory field $x_{i j}$ (restricted so that $x_{i j} \in[0, 1]$) with its own dynamics, giving (see cartoon in Fig. \ref{fig:spin_glass_viz})

\begin{equation}
\dot{\bf x}={\bf F}'({\bf x})\equiv\left\{
\begin{split}
    \dot{s}_i &= \sum_{<i j>} J_{i j} s_j - g \sum_{<i j>} x_{i j} s_i\,, \\
    \dot{x}_{i j} &= \gamma \big( C_{i j} - \delta \big)\,. \label{eq:memory_dynamics}
\end{split}
\right
.
\end{equation} 

Above, $C_{i j} \equiv (J_{i j} s_i s_j + 1)/2$, so $C_{i j} \in [0, 1]$. Furthermore, $g$ is the spin-memory coupling strength, and $\delta$ is a thresholding parameter for the growth of $x_{i j}$. Both $g$ and $\delta$ are kept $\sim O(1)$ in all our simulations so $1/\gamma$ still approximately represents a memory DOF timescale. These dynamics are similar (albeit simplified) to those from a previous study~\cite{PEI2022126727} in which memory is used as a tool to explore highly non-convex glassy landscapes. For the remainder of this work, we choose $g = 2$ and $\delta = 0.75$.

\begin{figure*}
    \centering
    \includegraphics[width=\textwidth]{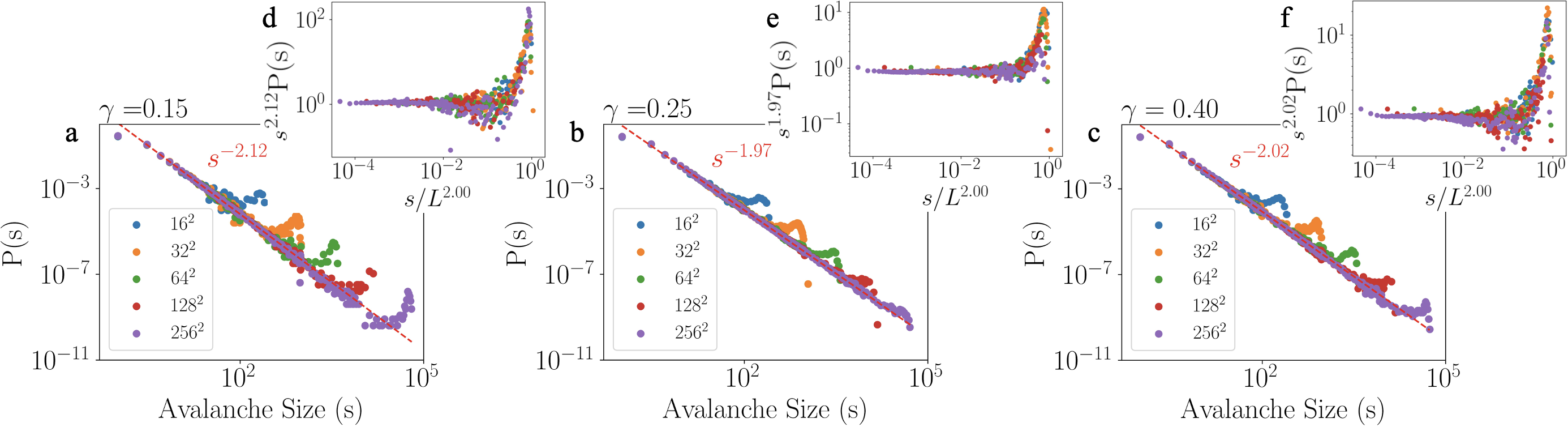}
    \caption{Avalanche size distributions (a-c) for the memory frequency scale $\gamma = 0.15$, $0.25$, and $0.40$ with scale-invariance analyses (d-f) performed in the upper-right insets. For each $\gamma$, distributions are plotted for lattice sizes $N \equiv L^2 = 16^2$, $32^2$, $64^2$, $128^2$, and $256^2$. We detect scale-free (SF) avalanches for all $\gamma \in \{0.15, 0.25, 0.40\}$, corroborated by the simultaneous collapse of all distributions in the scale-invariance analyses, with properly chosen exponents. Each lattice size is simulated in 100 distinct instances over a time $T = 200$. See the SM for more details. \label{fig:SF_phase}} 
\end{figure*}

We can now simulate \cite{github_link} these dynamics for a wide variety of $\gamma$, the memory DOF frequency scale. Spin and memory dynamics for a randomly chosen site $i$ in a small ($N = 16^2$) lattice are shown in Fig. \ref{fig:spin_glass_viz}. See the SM for a more detailed discussion of this model's dynamics and our chosen parameters.

\section{Avalanche distributions and phase structure}
To characterize the degree to which effective long-range interactions are present in these dynamics, we study avalanche distributions. We define an avalanche as a cluster of orthogonally connected spin flips that occur in recent succession (see the SM for a more detailed definition). Heuristically, these can be thought of as ``waves'' of spin flips that propagate through the lattice.

In Fig. \ref{fig:SF_phase}, we plot avalanche size distributions for $\gamma = 0.15$, $0.25$, and $0.40$ for lattice sizes $N \equiv L^2 = 16^2$, $32^2$, $64^2$, $128^2$, and $256^2$. Interestingly, in this range of $\gamma$, these distributions are all fitted well by power-law decays with exponents $\alpha_\gamma \sim 2$ (see also discussion below).

The fact that these avalanche distributions are scale-free (SF) suggests that, for $\gamma \lesssim 1/2$, LRO is present. Large avalanches correspond to instances during the system's dynamics when a large portion of spins in the lattice flip in quick succession, indicating that these spins, which are spatially distant for sufficiently large avalanches, become strongly correlated due to the presence of memory. The ``bumps'' in Fig. \ref{fig:SF_phase}, occurring at sizes just smaller than $N$ for a given lattice, correspond to an oversampling of {\it system-wide} avalanches~\cite{Pruessner_2012}. This is evident from our scale-invariance analysis, in which all bumps collapse when normalized by $N \equiv L^2$. Thus, as anticipated in our previous analysis, memory can induce strong coupling {\it throughout the entire lattice}. As we consider larger and larger lattices, these bumps shift to the right, and the range of power-law behavior increases accordingly. This indicates that true SF behavior likely persists in the thermodynamic limit.

For $\gamma < 0.15$, we also observe SF behavior, although our limited computational resources prevent us from simulating the largest ($N=256^2$) lattices (in the SM, we find SF avalanche distributions for $\gamma = 10^{-2}$ up to $N = 128^2$). Because SF distributions exist over a wide range of $\gamma$, we recognize this regime as a LRO {\it phase}, rather than a critical point of transition between two different phases.

We must emphasize that this behavior cannot be attributed to the theory of self-organized criticality (SOC) \cite{bak1987self}. This theory suggests that a system may {\it self-organize} to a critical point, yielding scale-free behavior. However, this framework requires there to be some conserved quantity (energy, resources, etc.) to achieve this ``critical-like'' behavior \cite{bonachela2009self}, which does not exist in our model. Formally, it also requires infinite timescale separation, which is not at all the case in our model. Furthermore, the extreme width of our phase of memory-induced LRO (appearing to have no lower bound) suggests to us this cannot be the result of the system ``tuning'' itself towards some truly critical state.

For $\gamma \gtrsim 1/2$, the dynamics are qualitatively different. In this regime, the memory is able to respond to changes in the spins much more rapidly, so quickly that the memory decays before some spins are able to flip. This process is cyclic, giving rise to quasiperiodic $s_i(t)$ orbits near $s_i = \pm 1$. So, avalanches are far less common in this regime. In the SM, we discuss this further and show that $s_i(t)$ is oscillatory in the limit of $\gamma \gg 1$. Lastly, we provide evidence that the LRO phase is {\it attractive} in the SM.

\section{Long-range correlated percolation}
To gain deeper insights into the observed LRO phase, we examine the problem through the lens of a correlated percolation transition.

\begin{figure*}
    \centering
    \includegraphics[width=0.95\textwidth]{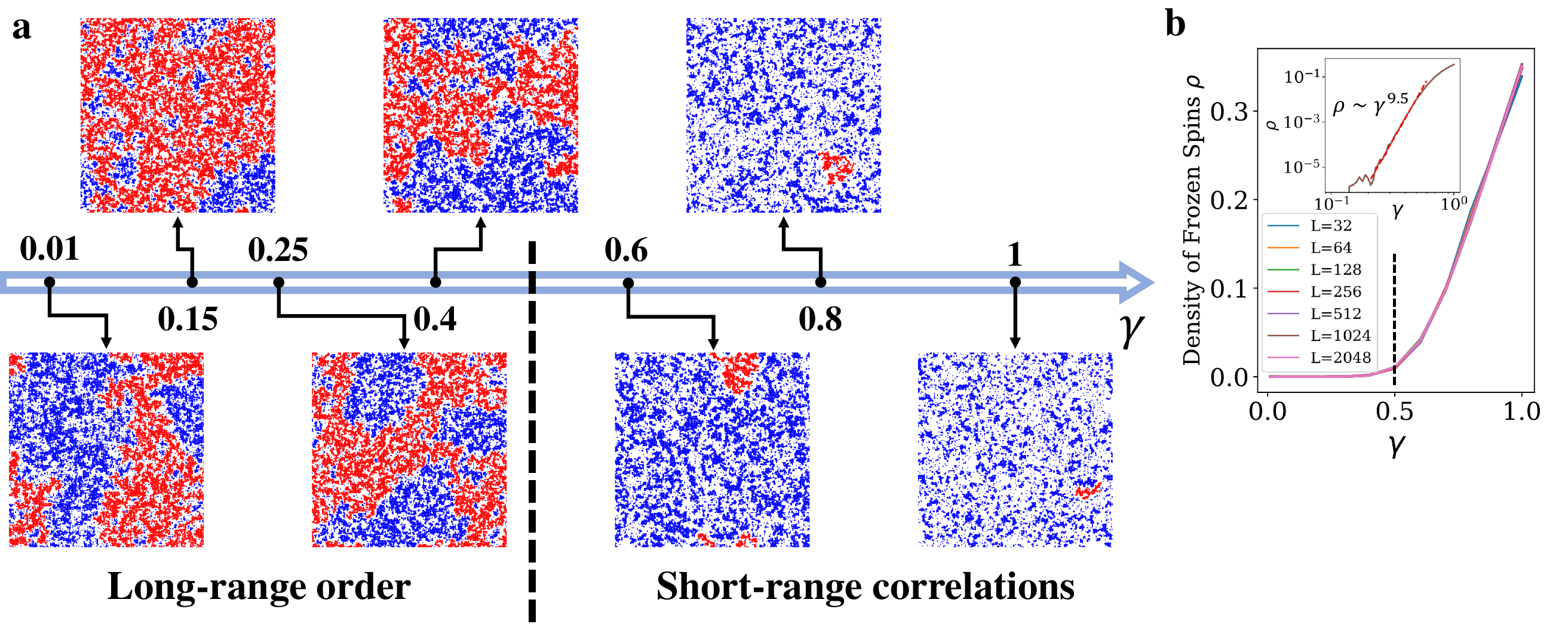}
    \caption{(a) Visualization of the transition into the LRO phase as a type of dynamic percolation on a $256\times 256$ lattice. In each snapshot, sites are considered occupied (colored) if their corresponding spin flips within $5 \times 10^3$ time steps, where each time step $\Delta t = 0.048\gamma^{-1/3}$. The largest connected cluster is highlighted in red (with periodic boundary conditions), while all smaller clusters are shown in blue. Only in the LRO phase do percolating clusters exist. A visualization of how the percolating clusters dynamically build up can be found in the Supplemental Movies. (b) The density of frozen spins, $\rho$, as a function of $\gamma$. For small $\gamma$, no spins are frozen, and the system is in the percolating phase. As $\gamma$ increases, more spins become locked into quasiperiodic orbits and no longer flip, driving the system toward a phase with short-range correlations and no system-wide clusters. The phase transition is continuous, and $\rho(\gamma)$ is independent of system size $L$. The inset shows $\rho(\gamma)$ in log-log scale. Up to the transition point $\gamma \simeq 0.5$, $\rho(\gamma)$ follows an approximate power-law relationship, $\rho\sim\gamma^{9.5}$.
    \label{fig:percolation}} 
\end{figure*}

First, we must define the percolation problem associated with our spin system with memory. To do this, we compare the spin configuration $\mathbf{s}(0)$ (after some transient time has already elapsed) to $\mathbf{s}(t)$, after a further time $t$ has passed. We consider a site ``occupied'' if the corresponding spin flips between $\mathbf{s}(0)$ and $\mathbf{s}(t)$; otherwise, the site is unoccupied (more details in the SM). Snapshots for various values of $\gamma$ are shown in Fig.~\ref{fig:percolation}. Interestingly, the transition into the LRO phase can also be visualized as a percolation transition: when $\gamma \lesssim 0.5$, the largest connected cluster (red) spans the entire system, whereas no spanning cluster exists for $\gamma \gtrsim 0.5$.

Things become more intriguing when we examine our system near the percolation threshold (see analysis in the SM). The occupation probability $p$ was computed as the average probability of a spin flipping after time $t$. In the limit $t \to \infty$, we find $p \to 0.5$. We calculate the percolation probability $\Pi(p, L)$, the percolation strength $P(p, L)$, the average cluster size $S(p, L)$, and perform a finite-size scaling analysis. This analysis yields a percolation threshold of $p_c = 0.5134 \pm 0.0005$, along with critical exponents $\nu_c = 1.852 \pm 0.001$, $\beta_c = 0.131 \pm 0.006$, $\gamma_c = 3.10 \pm 0.01$, and fractal dimension for the infinite cluster $D = 1.924 \pm 0.001$. These values deviate significantly from those of 2D uncorrelated site percolation on a square lattice, where $p_c = 0.5927$, and share some similarities with (2+1)D (2 spatial + 1 temporal dimensions) percolation (see SM). However, since our system is both dynamical and has long-range spin correlations, it falls firmly into its own universality class.

We found that our system's behavior aligns more closely with the phenomenon of long-range correlated percolation \cite{prakash1992structural}, where the spatial correlation of occupancy variables follows a power law: $\langle s(\mathbf{r}) s(\mathbf{r+R}) \rangle \sim R^{-a}$. Previous studies \cite{prakash1992structural} have shown that the percolation threshold $p_c$ decreases compared to the uncorrelated case as the exponent $a$ decreases, and in the limit $a \to 0$, $p_c \to 0.5$. Our observed $p_c = 0.5134$ (much less than 0.5927) thus corroborates the evidence that strong, long-range spin correlations exist. Also, the fact that $p \to 0.5$ as in the long-time limit suggests that our system is attracted to a point near the percolation transition.

With the established percolation analogy, we can describe the phase transition between long-range order and short-range correlations in greater detail. As discussed, when $\gamma \gtrsim 0.5$, some spins become locked into quasiperiodic orbits near $s_i = \pm 1$ and can no longer flip. We refer to these as ``frozen'' spins and denote their density as $\rho$. Fig.~\ref{fig:percolation}(b) shows $\rho$ as a function of $\gamma$. For small $\gamma$, no spins are frozen, and the system remains in the percolating LRO phase. As $\gamma$ increases, more spins become locked into quasiperiodic orbits, pushing the system toward an absorbing phase with no system-wide clusters. This eliminates some of the paths between spins, reducing the memory's effectiveness at inducing LRO. It is important to note that this phase transition is continuous: $\rho$ is independent of system size $L$. Up to the transition point $\gamma \simeq 0.5$, $\rho(\gamma)$ follows an approximate power-law relationship without discontinuities.


\section{Conclusions}
We have shown, both by employing general analysis and numerical simulations, that time non-locality (memory) can induce LRO in dynamical systems. The ensuing {\it phase} occurs when the memory DOFs are slower than the primary DOFs. This phase is non-perturbative and not related to criticality.

We expect our analysis to generalize to a wide variety of dynamical systems (since it is largely independent of the precise functional form taken by the primary and memory DOF dynamics). Recent experimental and numerical work has already shown that memory can induce a phase of long-range order in arrays of thermal neuristors, a new type of spiking neuron \cite{qiu2024reconfigurable, yuanhang_ivan_thermal_neuristors}, and biologically-inspired neural arrays \cite{sun2024memoryneuralactivitylongrange, morrell2021latent}. We also expect that this phenomenon could be detected in recently developed electronic systems, such as single-layer MoS$_2$ transistor \cite{mos2_transistors} and neuromorphic spintronic \cite{neuromorphic_spintronics} arrays. In fact, we suspect that a wide class of dynamical systems may exhibit this behavior, so long as they feature at least two types of dynamical variables which evolve according to different characteristic timescales.

Regarding the usefulness of this phenomenon from a more practical perspective, a new computing paradigm known as Memcomputing \cite{di2018perspective, di2022memcomputing, sipling2025phase} is already leveraging this phenomenon to solve difficult combinatorial optimization problems efficiently. In these machines, the original, logical variables act as the primary DOFs, and auxiliary memory DOFs are added to assist in the solution search process. The emergence of LRO enables these devices to evolve {\it collectively}, providing an advantage over traditional algorithmic approaches.

Finally, although challenging, for future work, it would be interesting to perform a Renormalization Group (RG) analysis on these types of systems.

{\it Acknowledgements}.---C.S., Y.H.Z., and M. D. are supported by NSF grant No. ECCS-2229880.

\bibliography{apssamp}

\end{document}


\title{Supplemental Material for ``Memory-induced long-range order in dynamical systems''}

\author{Chesson Sipling}
\email{email: csipling@ucsd.edu}
\affiliation{Department of Physics, University of California San Diego, La Jolla, CA 92093}

\author{Yuan-Hang Zhang}
\email{email: yuz092@ucsd.edu}
\affiliation{Department of Physics, University of California San Diego, La Jolla, CA 92093}

\author{Massimiliano Di Ventra}
\email{email: diventra@physics.ucsd.edu}
\affiliation{Department of Physics, University of California San Diego, La Jolla, CA 92093}

\maketitle

\section{Model Details}

For clarity, we rewrite the coupled spin-memory dynamics in the main text here:

\begin{equation}
\left\{
\begin{split}
    \dot{s}_i &= \sum_{<i j>} J_{i j} s_j - g \sum_{<i j>} x_{i j} s_i\,, \\
    \dot{x}_{i j} &= \gamma \big( C_{i j} - \delta \big) \label{eq:memory_dynamics}\,.
\end{split}
\right
.\end{equation}

According to these dynamics, each spin $s_i$ is encouraged to align or anti-align with its neighbors $s_j$ depending on the sign of $J_{i j}$.

The ``satisfaction function'' $C_{i j} \equiv (J_{i j} s_i s_j + 1)/2$ quantifies the degree to which a pair of spins $s_i$ and $s_j$ satisfy their $J_{i j}$ interaction. For example, $C_{i j} = 0$ corresponds to $s_i s_j = -J_{i j}$, so that the spins are anti-aligned if $J_{i j} = 1$ and aligned if $J_{i j} = -1$. This would be undesirable from an energy-minimization perspective for a true spin glass (without any memory coupling or dynamics). $C_{i j} = 1$ instead corresponds to perfect alignment/anti-alignment {\it in accordance} with $J_{i j}$.

\subsection{Choice of Memory Dynamics} 

We chose the form for the memory dynamics in Eq.~(\ref{eq:memory_dynamics}) since it is one of the conceptually simplest ways to introduce the $s_i s_j$ coupling present in the general analysis of the main text, without being so simple that the dynamics prematurely stop. Due to the high non-convexity of the relaxed spins' configuration space, the bare spin system would quickly get ``stuck'' at a point in configuration space (which is most likely {\it not} the global minimum) were it not for the coupled memory dynamics. By having $\dot{x}_{i j}$ depend on $C_{i j}$, $x_{i j}$ will only grow when a spin-spin interaction is sufficiently ``satisfied'' ($C_{i j} > \delta$). This, in turn, increases the effect of the spin-memory coupling in the spin dynamics, ``pushing'' $s_i$ away from its value which is in accordance with $J_{i j}$ and $s_j$, and preventing the dynamics from stopping. So, typically, the memory dynamics decay when a spin is flipping and grow once a spin has stabilized at $\pm 1$.

To further clarify this, consider the effect the memories have on the spins (and vice versa) during typical system dynamics. During much of their dynamics, individual spins take values near $\pm 1$, occasionally flipping over a relatively short timescale. While flipping, since most spins are moving away from a state that was desirable to the spin-spin interaction $\sum_{<i j>} J_{i j} s_j$, $C_{i j}$ will often decrease, causing the memory to decay. On the other hand, at sites $i$ where spins have stabilized at $\pm 1$, the memories $x_{i j}$ will often grow. This will encourage later flips, preventing the dynamics from halting.

\subsection{Choice of Parameters}

We will now clarify why we choose $g = 2$ and $\delta = 0.75$ when simulating these dynamics. The parameter $g = 2$ is chosen since each $x_{i j}$ has an average value of around 1/2 during its dynamics (as also observed during our simulations). Since $|J_{i j}| = 1$, $g = 2$ makes $|J_{i j} s_j|$ and $|g x_{i j} s_i|$ most comparable, so neither term dominates the spin dynamics. Note that in the limit where $g \rightarrow 0$, the memories decouple from the spins, and the spin dynamics reduce to that of a traditional (continuously relaxed) spin glass. On the other hand, $g \rightarrow \infty$ renders the $J_{i j}$ coupling term negligible, causing all spins to decay exponentially towards 0 (since $x_{i j} \geq 0$).

The parameter $\delta = 0.75$ was chosen to provide a balance between two extremes: $\delta \approx 1/2$ and $\delta \approx 1$. $\delta \approx 1/2$ encourages frequent spin oscillations around 0, so $|s_i| \approx 1$ becomes rare. Ideally, we would like this system to still resemble a spin glass, with $s_i$ remaining near $\pm 1$ during most of their dynamics, punctuated by sudden flips due to the presence of the dynamic memory field that prevents the spin dynamics from getting stuck prematurely. This makes $\delta > 1/2$ desirable. Alternatively, $\delta \approx 1$  would require a great deal of computational resources to extract statistically significant spin-flip data, as $x_{i j}$ would only grow when an interaction was ``fully'' satisfied ($C_{i j} = 1$). 

In the especially interesting regime where the memory dynamics are slow, if we had $\delta \approx 1$, it would take a long time between bouts where $x_{i j}$ is large enough to encourage spins to flip. Note that we must have $\delta \in (1/2, 1)$ since it acts as a thresholding variable for $C_{i j}$, which only indicates spin-$J_{i j}$ {\it satisfaction} for $C_{i j} \in (1/2, 1)$. Furthermore, if we had $\delta \geq 1$, all $x_{i j}$ would quickly shrink to 0, decoupling the memory field from the spin dynamics entirely. We expect our results to hold (i.e., that the LRO phase would persist) even if $g$ and $\delta$ were perturbed from our chosen values.

Additionally, for each simulated instance, we randomly initialize $s_i(0)$ and $J_{i j}$ to $\pm 1$ (with $J_{i j}$ symmetry enforced), setting all $x_{i j}(0) = 0.5$.

\subsection{Quasiperiodic $s_i(t)$ Orbits in the $\gamma \gg 1$ Limit}

In the main text, we briefly mention why we expect some spins to follow quasiperiodic orbits in the $\gamma \gtrsim 1/2$ limit, yielding fewer avalanches. A more in-depth explanation will now be given.

\begin{figure}
    \includegraphics[width=\textwidth]{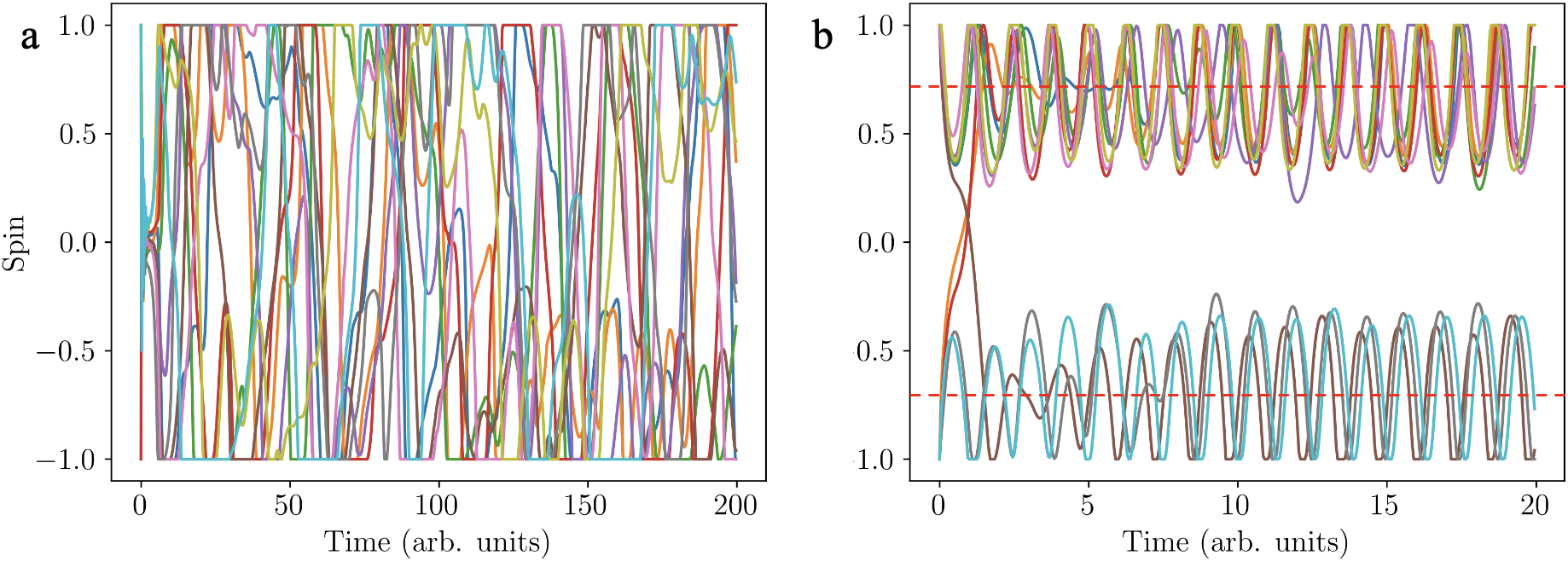}
    \caption{Spin dynamics for 10 randomly chosen spins in a square lattice of size $N = 16^2$ with $\gamma = 0.2$ (a) and $\gamma = 10$ (b). In the $\gamma = 10$ case, spin flips are almost entirely absent following an initial transient. Most spins follow quasi-periodic orbits around $\pm \sqrt{2 \delta - 1} = 1/\sqrt{2} \approx 0.707$ (for $\delta = 0.75$), marked by red horizontal lines. \label{fig:spin_dynamics}}
\end{figure}

Consider a single frustrated spin $s_i = \pm 1$. For that spin to move away from $\pm 1$, the spin-memory coupling $g \sum_{<i j>} x_{i j} s_i$ must exceed the spin-spin coupling $\sum_{<i j>} J_{i j} s_j$. However, this will also decrease $C_{i j}$, and, for sufficiently large $\gamma$, this can induce rapid memory decay. In many cases, the memory decays so quickly that the spin-spin coupling dominates once again before $s_i$ has changed sign. $|s_i|$ then increases back towards its original, frustrated value. We can see that this process will be cyclic, giving rise to quasiperiodic $s_i(t)$ orbits near $s_i = \pm 1$. So, avalanches are far less common in this regime. Fig.~\ref{fig:spin_dynamics} shows how large-$\gamma$ ($\gamma = 10$) spin dynamics differ from a case where we observe many more spin flips ($\gamma = 0.2$).

For $\gamma \gg 1$, this effect becomes even more pronounced. We expect the amplitude of spin oscillations to decrease, as the spins have even less time to respond to sudden changes in the memories: $x_{i j}$ will rapidly increase or decrease depending on whether $C_{i j} > \delta$ or $C_{i j} < \delta$. The marginal case, $C_{i j} = \delta$, constrains the equilibrium values around which $s_i$ oscillate. Requiring that $\delta > 1/2$ (i.e., $s_i$ and $s_j$ are always {\it in accordance} with the $J_{i j}$ interaction when $C_{i j} > \delta$), we can make the approximation $|s^{eq}_i| = |s^{eq}_j|$ and solve for the $s^{eq}_i$ that gives $C_{i j} = \delta$:

\begin{equation}
\begin{split}
    \frac{1}{2}(|s^{eq}_i|^2 + 1) = \delta \implies s^{eq}_i = \pm \sqrt{2 \delta - 1}\,.
\end{split}
\end{equation}

The spin dynamics in Fig.~\ref{fig:spin_dynamics}(b) indeed appear to oscillate around these $s^{eq}_i$. This suggests that, for large enough $\gamma$, nearest-neighbor spins ``lock in'' to quasiperiodic orbits around $\pm \sqrt{2\delta - 1}$, with the sign determined by the nature of their mutual $J_{i j}$ interaction (ferromagnetic or antiferromagnetic).

\section{Avalanche Definition, Binning, and Extraction}

As mentioned in the main text, we define an avalanche as a collection of spin flips that occur in recent succession. Here, we provide a few more details regarding how avalanches are defined, extracted, and binned in this work.

More formally, we define an avalanche as follows. Imagine that a spin at an arbitrary lattice site $i$ flips. Wait for a time $\Delta_{tw}$ (in units of the fixed numerical timestep), which we call the {\it time window}. During this window, the spin $s_i$ is considered ``active''. If one of its nearest neighbors $s_j$ flips before a time $\Delta_{tw}$ passes, $s_j$ joins the avalanche, increasing the avalanche's size by $1$. $s_j$ is then active for a duration $\Delta_{tw}$, starting from when it first flipped, during which time the avalanche can become even larger if one of its nearest-neighboring spins, $s_k$, flips. Note that $s_i$ is still considered active until a time $\Delta_{tw}$ passes from when it originally flipped. This enables the avalanche to propagate outwards in a variety of different directions.

If none of the spins adjacent to a particular spin, $s_a$, flip after a time $\Delta_{tw}$, $s_a$ becomes ``inactive'' (subscript $a$ indicating an arbitrary active site in an existing avalanche). Once all spins in an avalanche become inactive, the avalanche terminates, and the total number of sites at which spins flipped defines that avalanche's size. In the case where the first spin in the avalanche, $s_i$, becomes inactive before any of its nearest neighbors flip, we record an avalanche of size $1$.

Notably, we do not allow spins that previously flipped in a given avalanche to count ``double'' toward an avalanche's size (although they can allow an avalanche to continue propagating). This assures that the maximum possible avalanche size is $N$. Furthermore, note that many avalanches can exist in a lattice at any instant in time. If two or more avalanches meet (a spin flips which is a nearest-neighbor of two or more active spins in different avalanches), all involved avalanches are combined. Their sizes are added to determine the size of the new, composite avalanche.

Before plotting these avalanche distributions, we bin avalanches logarithmically with bins of size $2$ for $s < 10^2$, bins of size $20$ for $10^2 \leq s < 10^3$, bins of size $200$ for $10^3 \leq s < 10^4$, etc. Logarithmic binning is generally accepted as more reliable for data that spans many orders of magnitude \cite{Pruessner_2012, doi:10.1137/070710111}. We use this precise binning rather than an evenly spaced one because it gave the best agreement in scale-free exponent $\alpha$ between the raw and binned data (in the regime in which scale-free behavior is observed). For example, for $\gamma = 0.25$, $N = 256^2$, and $\Delta_{tw} = 5$, the pre-binning $\alpha = 1.92$ and post-binning $\alpha = 1.97$ agree fairly well. Other logarithmic binning approaches gave differences of 5\% or more or didn't fit to a single scale-free exponent at all. We use $\Delta_{tw} = 5$ for all distributions in the main text. These particular $\Delta_{tw}$ choices are explained further (through the lens of percolation) in the ``Long-Range Correlated Percolation'' section below.

Lastly, the first data point from each avalanche distribution is omitted from all scale-invariance analysis plots (since a scale-free distribution cannot be normalized, we do not expect power-law behavior below some $s_{min}$~\cite{doi:10.1137/070710111}).

\section{Phase Structure}

\subsection{Persistence of LRO Phase for $\gamma < 0.15$}

\begin{figure*}
    \centering
    \includegraphics[width=0.5\textwidth]{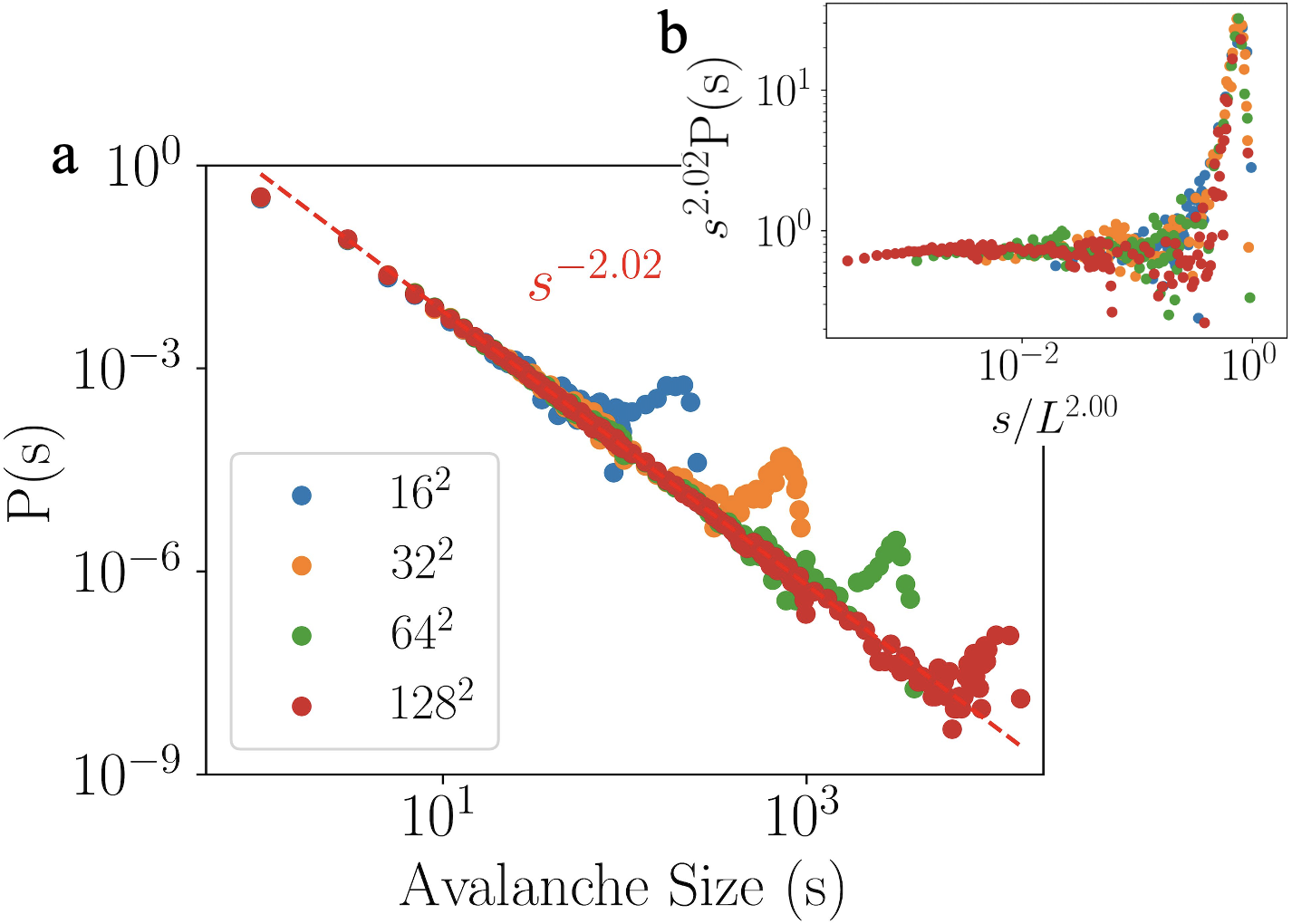}
    \caption{Avalanche size distributions (a) for $\gamma = 10^{-2}$ and lattice sizes $N \equiv L^2 = 16^2$, $32^2$, $64^2$, $128^2$ with scale-invariance analysis (b) performed in the upper-right inset. We detect scale-free avalanches and a similarly good finite-size scaling analysis as for $\gamma \in [0.15, 0.40]$. Each lattice size is simulated in 100 distinct instances, each of which is run for a time $T = 1000$. Avalanches are captured using a time window of $\Delta_{tw} = 16$. \label{fig:small_gamma_LRO}}
\end{figure*}

We plot avalanche size distributions for $\gamma = 10^{-2}$ and $N = 16^2$, $32^2$, $64^2$, and $128^2$ in Fig.~\ref{fig:small_gamma_LRO}. Because a lower memory frequency $\gamma$ extends the time between which we typically observe spin flips, we must simulate these instances up to a time $T = 1000$ rather than $T = 200$ (used for $\gamma \in [0.15, 0.40]$) to get comparable statistics. The reasonably good power law fit with an exponent $\alpha_{0.01} \sim$ 2 indicates that the memory still effectively couples spins at long distances for low $\gamma$. Additionally, the characteristic bumps continue to shift, as expected, with increasing $N$. Although our computational limitations prevent us from simulating even smaller $\gamma$, these results suggest that the LRO phase likely persists for arbitrarily small $\gamma$.

\subsection{Avalanche Distributions for $\gamma \gtrsim 1$}

As mentioned in the main text and further detailed in a previous SM section, far fewer spins flip when $\gamma \gtrsim 1$. Inevitably, this yields fewer and smaller avalanches. However, given a sufficiently large sample size, we can still extract some meaningful information about the distribution of avalanches that {\it do} occur.

\begin{figure*}
    \centering
    \includegraphics[width=0.9\textwidth]{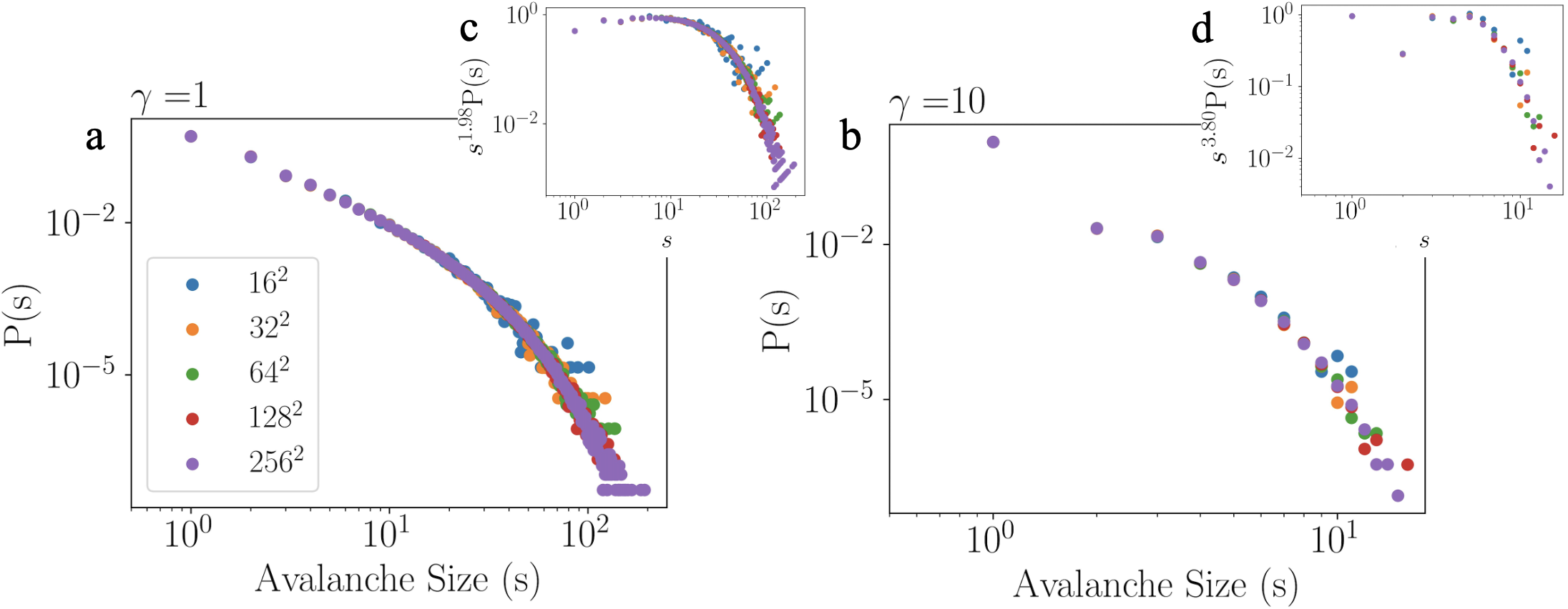}
    \caption{Avalanche size distributions for $\gamma = 1$ (a) and $\gamma = 10$ (b), for lattice sizes $N \equiv L^2 = 16^2$, $32^2$, $64^2$, $128^2$, and $256^2$, with scale-invariance analysis (c-d) performed in the upper-right insets (with scale-free exponents $\alpha_\gamma$ fitted manually in this case). The lack of larger avalanches as $\gamma$ increases suggests that the effective range of interactions shortens as the memory timescale shortens. Each lattice size is simulated in 250 distinct instances, each of which is run for a time $T = 100$. Avalanches are captured using a time window of $\Delta_{tw} = 2$. No binning is performed; the raw data is plotted. The data wasn't fitted well by either Gaussian or scale-free distributions. \label{fig:short_range_ints}}
\end{figure*}

In Fig.~\ref{fig:short_range_ints}, we plot avalanche distributions for $\gamma = 1$ and $\gamma = 10$, each with lattice sizes $N = 16^2$, $32^2$, $64^2$, $128^2$, and $256^2$, for 250 instances. Evidently, smaller avalanches are the norm when the memory timescale is comparable to or even shorter than the spins' dynamic timescale. Furthermore, this distribution decays quite rapidly and cannot be fitted well to either a Gaussian or a scale-free distribution $P(s) \sim s^{-\alpha}$ (although we do find $\alpha$ manually which gives the ``best'' scale-invariance analysis). This suggests that the effective range of interactions shortens significantly when $\gamma$ is too large, in agreement with our previous analysis.

\subsection{Attractiveness of the LRO Phase}

Now, we highlight that the LRO phase is {\it attractive}. This means that the system is encouraged to enter the LRO phase as time goes on. This is reminiscent of the phenomenon of ``self-organized criticality''~\cite{Pruessner_2012}. However, as already mentioned, the effect we describe is {\it not} associated with criticality since the dynamics are not attracted to any particular point in the phase diagram. 

\begin{figure}
    \centering
    \includegraphics[width=0.5\textwidth]{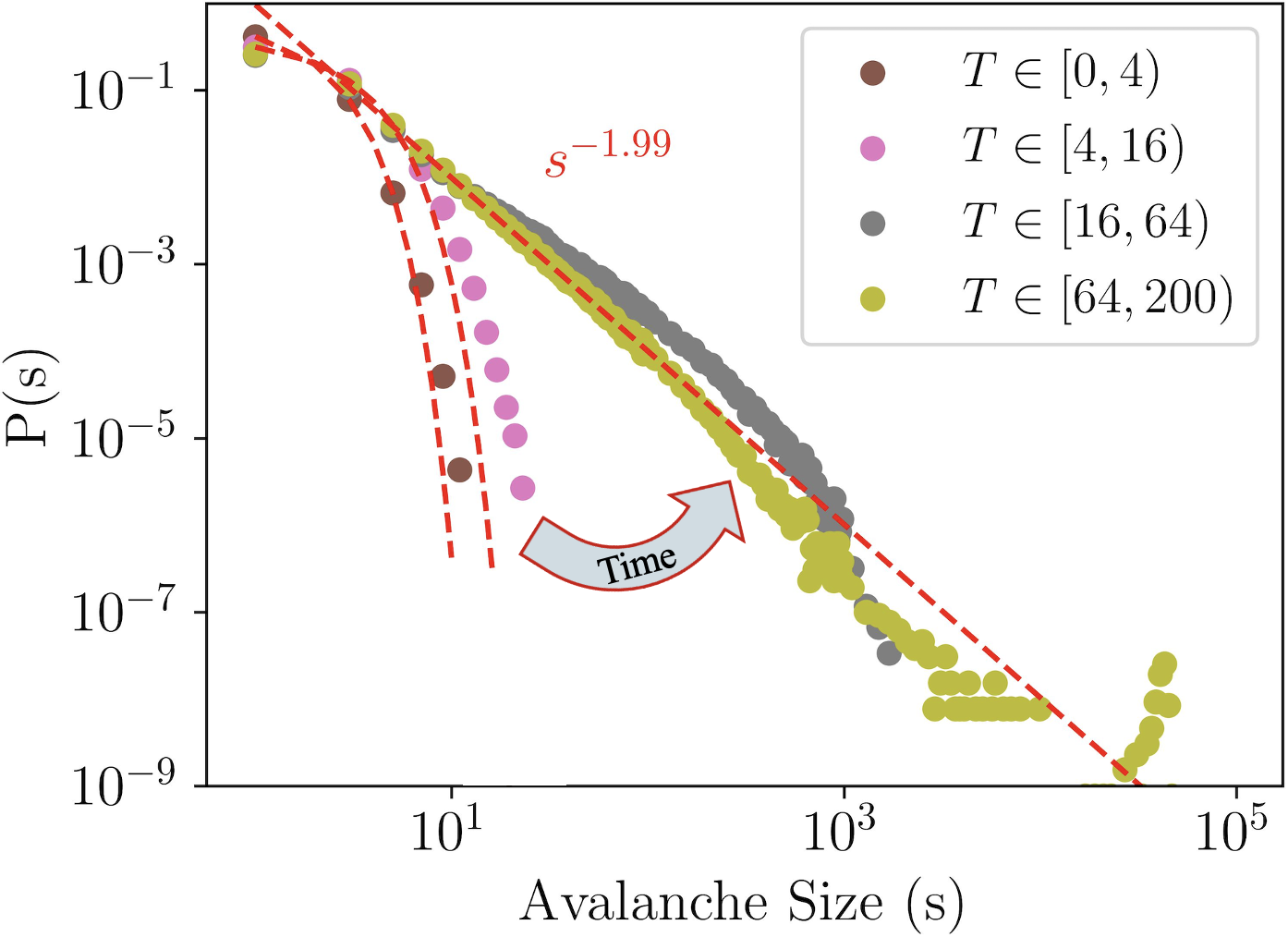}
    \caption{Avalanche size distributions for $\gamma = 0.25$, $N = 256^2$ across 4 time intervals. As time goes on, the distributions approach scale-free (SF). The first two intervals are fitted to Gaussian distributions while the last is fitted to an SF distribution. 100 instances are simulated over a time $T = 200$. \label{fig:attractive_LRO}}
\end{figure}

Recall in our general analysis that we began to first recognize a dependence on sites $k$, the next-nearest-neighbors of $i$, after a time $T \gtrsim 2/\gamma$. Furthermore, higher-ordered couplings between further sites only emerge at later times. Therefore, some finite time must pass before the spins become correlated at long distances. To study this quantitatively, we plot avalanche size distributions for $\gamma = 0.25$, $N = 256^2$ over time intervals $T \in [0, 4)$, $T \in [4, 16)$, $T \in [16, 64)$, and $T \in [64, 200)$ in Fig. \ref{fig:attractive_LRO}. During earlier time intervals, the distributions are nearly fitted by Gaussians, although deviations from pure Gaussians are present. This departure is more pronounced for the later intervals, as beyond-nearest-neighbor interactions begin to emerge, and the distributions approach SF.

The interval $T \in [16, 64)$ seems to oversample intermediate-size avalanches, ``overshooting'' the SF distribution approached for the last case, $T \in [64, 200)$. This is artificial: many avalanches that may have become much larger are ``cut off'' when we truncate the simulation at $T = 64$. This is analogous to how avalanche sizes near $N$ are oversampled due to the finite spatial extent of a simulated lattice.

\section{Memory Correlations}

As further evidence that the spins in our model \ref{eq:memory_dynamics} exhibit LRO, we study correlations between the memory DOFs. This is useful because the memories can be thought of as constituting an effective (dynamic) potential that influences the spins. Thus, if the correlations between memory DOFs do not decay too quickly, we would also expect that the effective spin-spin interaction generated due to the presence of memory is long-range.

\subsection{Memories as ``Spins''}

However, similarly to the spins, the correlations between the memory DOFs alone do not tell the whole story. For the spin DOFs, we  studied the correlations {\it indirectly} by analyzing avalanche distributions, which revealed that the spins can influence other spins to flip at long distances even though the orientations of spins as up/down may appear uncorrelated (since $J_{i j}$ are randomized in the lattice).

To study memory correlations, we will do something similar. We define a new DOF $\tilde{x}_{i j} \equiv \text{sgn}(\dot{x}_{i j})$, which keeps track of whether a memory DOF is increasing or decreasing. According to \ref{eq:memory_dynamics}, $\tilde{x}_{i j} > 0$ when $C_{i j} > \delta$ (the $J_{i j}$ interaction is ``satisfied'') and $\tilde{x}_{i j} < 0$ when $C_{i j} < \delta$ (the $J_{i j}$ interaction is ``unsatisfied''). This allows us to analyze avalanche distributions, constituted of $\tilde{x}_{i j}$ flips between $\pm 1$, in the same manner that we did for the spins.

\subsection{Memory Avalanche Distributions}

\begin{figure*}
    \centering
    \includegraphics[width=0.65\textwidth]{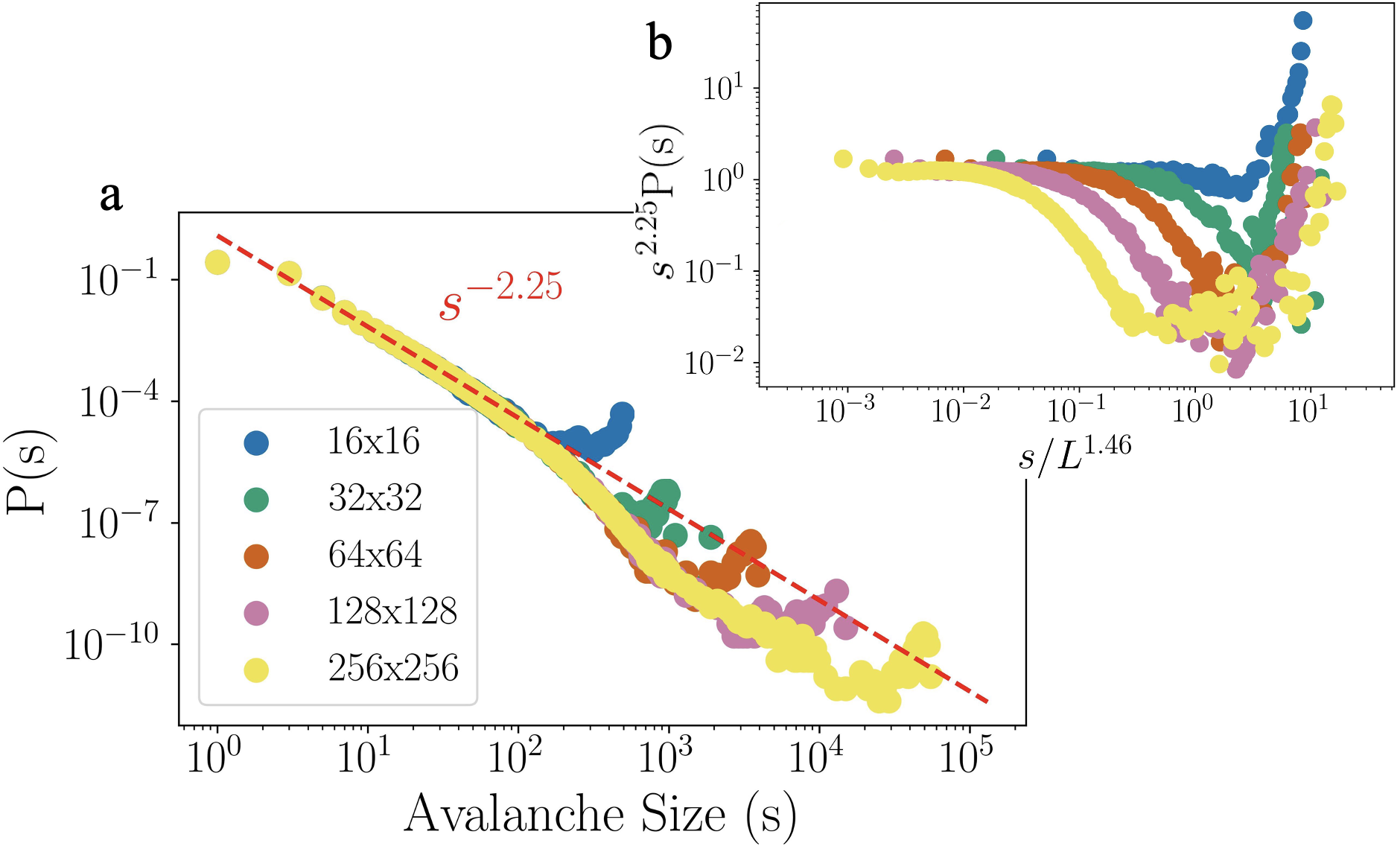}
    \caption{Memory avalanche size distributions (a) for $\gamma = 0.4$ and lattice sizes $N \equiv L^2 = 16^2$, $32^2$, $64^2$, $128^2$, and $256^2$ with scale-invariance analysis (b) performed in the upper-right inset. We detect long-range correlations but not scale-free behavior, as the distributions do not abide by the finite-size scaling ansatz. Each lattice size is simulated in 250 distinct instances, each of which is run for a time $T = 200$. Avalanches are captured using a time window of $\Delta_{tw} = 2$. \label{fig:mem_avalanche}}
\end{figure*}

In Fig.~\ref{fig:mem_avalanche}, we plot $\tilde{x}_{i j}$ avalanche distributions. Although these distributions do not quite satisfy the finite-size scaling ansatz, they decay polynomially to the lattice boundary. So, even though we would not refer to these distributions as ``scale-free'', they do show that the memories correlate strongly at long distances. In other words, the memory DOFs can be thought of as an effective long-range potential, further elucidating why the spins appear to interact at long range despite the lack of any explicit spin-spin coupling beyond nearest-neighbors.

\section{Long-range correlated percolation}

In this section, we delve deeper into the percolation transition associated with the spin dynamics.

\begin{figure*}
    \centering
    \includegraphics[width=\textwidth]{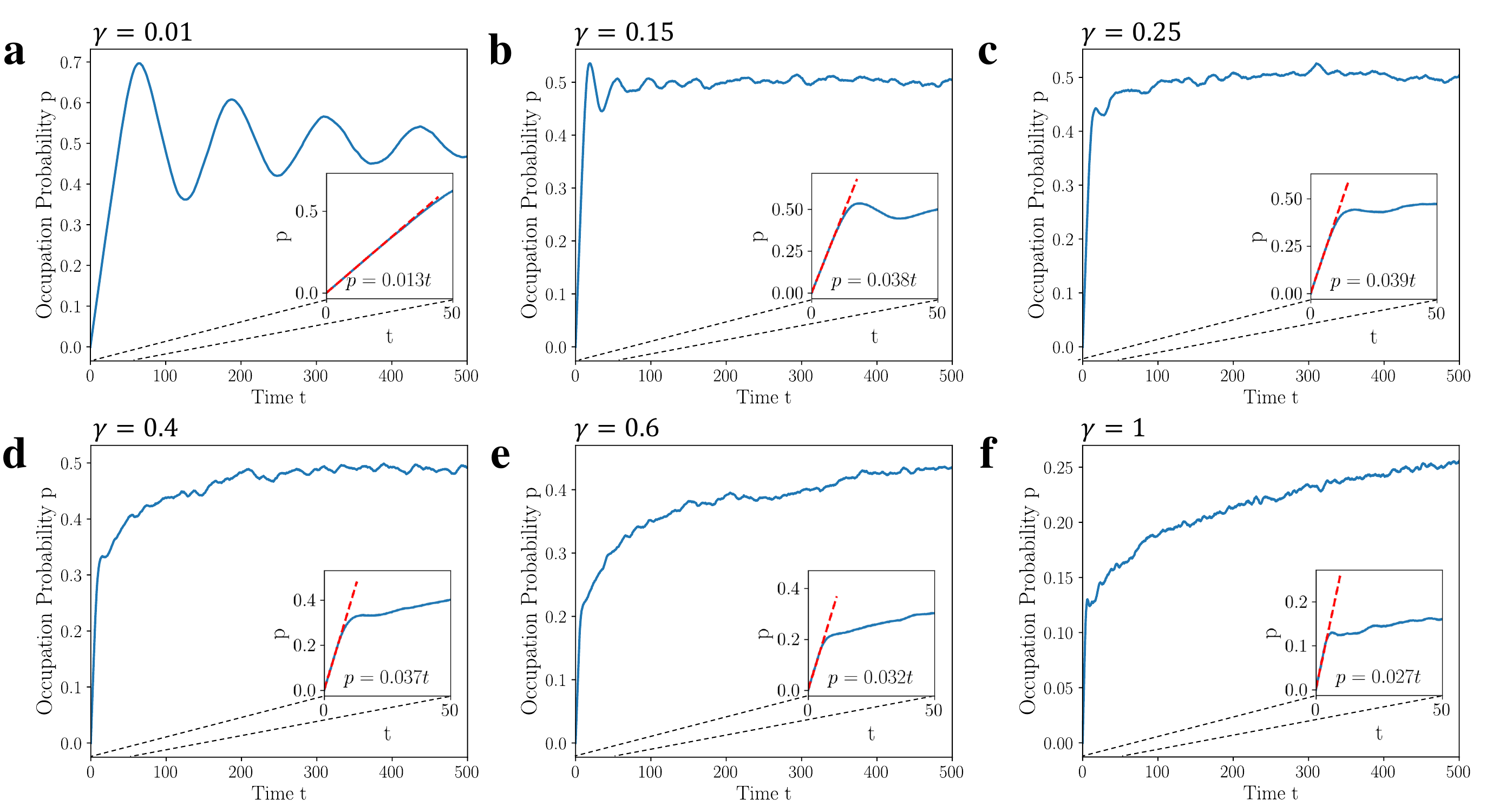}
    \caption{The occupation probability $p$ as a function of the time interval $t$ for various values of $\gamma$. The insets display linear fits of $p(t)$ near $t = 0$. For all $\gamma$ values, $p$ eventually converges toward $0.5$, but this convergence slows significantly as $\gamma$ increases due to an increasing number of spins becoming frozen in quasiperiodic orbits. The time window $\Delta_{tw}$, defined in the main text for avalanche extraction, satisfies $p(\Delta_{tw})=0.20$ for each $\gamma$, since we find $p_c = 0.20$ as an approximate percolation threshold for the 2+1D percolation defined on this system.\label{fig:p_vs_t}}
\end{figure*}

Recall that a site is defined as occupied if a spin flips within a time interval $t$, i.e., $s(0)s(t) < 0$. Thus, the occupation probability $p$ in the percolation problem is a function of time $t$ in the dynamical system. In Fig.~\ref{fig:p_vs_t}, we plot $p$ as a function of $t$ for various values of $\gamma$. For all $\gamma$ values, $p$ gradually converges toward $0.5$ as $t$ increases. For small $\gamma$, this convergence is rapid, accompanied by oscillatory behavior, which reflects strong memory effects. As $\gamma$ increases, some spins become locked into quasiperiodic orbits and cease to flip. Consequently, the occupation probability $p$ remains at smaller values for extended periods, keeping the system in the non-percolating phase and preventing the emergence of long-range order.

Next, we numerically determine the percolation threshold, $p_c$, and various critical exponents, through finite-size scaling analyses \cite{malthe2024percolation}. 

Near the percolation transition, the correlation length $\xi$ diverges as

\begin{equation}
    \xi\propto |p-p_c|^{-\nu}\Rightarrow |p-p_c|\propto \xi^{-\frac{1}{\nu}}\,.
\end{equation}

If a quantity $\Omega$ diverges near the percolation transition, it can be expressed as a function of $\xi$:

\begin{equation}
    \Omega\propto |p-p_c|^{-\omega}\propto \xi^{\frac{\omega}{\nu}}\,. \label{eq:X_xi}
\end{equation}

For a finite system with linear size $L$, when $\xi \ll L$, finite-size effects are negligible, and Eq.~\eqref{eq:X_xi} holds. However, when $\xi \gg L$, the correlation length is limited by the system size, leading to $\Omega \propto L^{\frac{\omega}{\nu}}$. This results in the scaling function

\begin{equation}
    \Omega(L, \xi)=\xi^{\frac{\omega}{\nu}}f(\frac{L}{\xi})\label{eq:finite}
\end{equation}
where
\begin{equation}
f(x)=\begin{cases}
\mathrm{const}, & x\to\infty, \\
x^{\omega/\nu}, & x\to 0.
\end{cases}
\end{equation}

Defining $g(x)=x^{-\omega}f(x^\nu)$, we can rewrite Eq.~\eqref{eq:finite} as
\begin{equation}
    \Omega(L, p)\sim L^{\omega/\nu} g(L^{1/\nu}|p-p_c|)\,. \label{eq:finite_size_scaling}
\end{equation}

By plotting $L^{-\omega/\nu}\Omega(L, p)$ against $L^{1/\nu}(p - p_c)$, we obtain the scaling function $g(x)$, which remains the same for different values of $L$. Thus, the correct values of $\omega$, $\nu$, and $p_c$ should collapse all curves for different $L$ onto the same scaling function. This property allows us to estimate the percolation threshold and critical exponents. 

\begin{figure*}
    \centering
    \includegraphics[width=\textwidth]{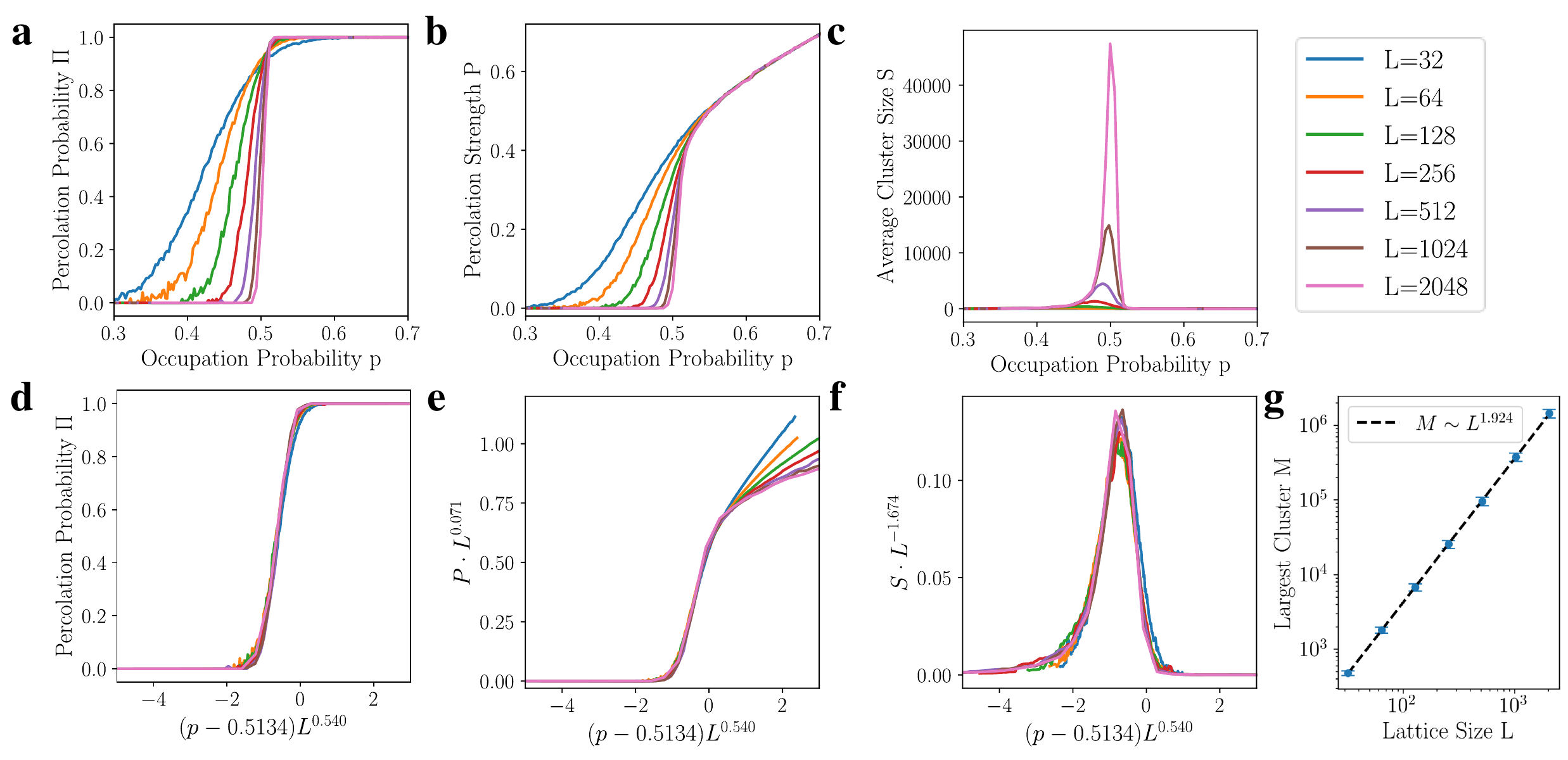}
    \caption{Finite-size scaling analysis of the percolation transition, with $\gamma=0.01$. (a) The percolation probability $\Pi(p, L)$. (b) The percolation strength $P(p, L)$. (c) The average cluster size (excluding the infinite cluster) $S(p, L)$. (d-f) Finite-size collapse of $\Pi(p, L)$, $P(p, L)$, and $S(p, L)$, respectively. When rescaled with the appropriate exponents, all curves for different system sizes $L$ collapse onto the same scaling function, demonstrating scale invariance. The exponents are optimized to achieve maximum overlap among the curves, yielding $p_c = 0.5134$, $1/\nu_c = 0.540$, $\beta_c/\nu_c = 0.071$, and $-\gamma_c/\nu_c = -1.674$. (g) The size of the infinite cluster $M$ as a function of $L$ at the percolation threshold $p_c = 0.5134$. The relationship $M \sim L^D$ defines the fractal dimension $D$ of the infinite cluster. A power-law fit gives $D = 1.924$.
    \label{fig:percolation_finite_size_scaling}}
\end{figure*}

We computed the percolation probability $\Pi$, percolation strength $P$, and average cluster size $S$ as functions of the occupation probability $p$ for system sizes $L = 32, 64, \dots, 2048$, with $\gamma=0.01$. The original and rescaled curves are plotted in Fig.~\ref{fig:percolation_finite_size_scaling}(a)-(f). Note that since the percolation probability $\Pi$ does not diverge, its corresponding exponent is zero; consequently, Fig.~\ref{fig:percolation_finite_size_scaling}(d) does not have a rescaled y-axis.

The percolation threshold and critical exponents were obtained by minimizing the standard deviation of a dense set of data points near the phase transition across different system sizes $L$, ensuring maximal overlap of the rescaled curves. Furthermore, the fractal dimension $D$ of the infinite cluster at the percolation transition was determined by fitting the size of the percolating cluster, $M$, as a function of the system size $L$. The values of the percolation threshold $p_c$, critical exponents $\beta_c$, $\gamma_c$, $\nu_c$, and the fractal dimension $D$ are summarized in Table~\ref{tab:exponents}.

In Table~\ref{tab:exponents}, we also compare the computed values of the percolation threshold and critical exponents with those for 2D uncorrelated site percolation \cite{christensen2002percolation} and long-range correlated percolation \cite{prakash1992structural}. Specifically, Ref.~\cite{prakash1992structural} studied 2D site percolation with a correlation function defined as

\begin{equation}
    \langle s(\mathbf{r})s(\mathbf{r}+\mathbf{R})\rangle =\int_{-\infty}^\infty |\mathbf{q}|^{-\lambda} e^{-i\mathbf{q}\cdot\mathbf{R}}d\mathbf{q}=f(\lambda)R^{-(d-\lambda)}. \label{eq:long_range_percolation}
\end{equation}

In Eq.~\eqref{eq:long_range_percolation}, the uncorrelated case corresponds to $\lambda = 0$, as $f(\lambda = 0) = 0$, while the correlation becomes stronger as $\lambda$ increases. Ref.~\cite{prakash1992structural} demonstrated that as $\lambda$ increases, $p_c$ decreases and approaches $p_c = 0.5$ as $\lambda \to 2$, while the correlation length exponent $\nu$ increases. The behaviors observed in our model qualitatively align with these trends (since our model's $p_c$ and $\nu_c$ are lower and higher, respectively, than the uncorrelated case), suggesting the presence of long-range correlations within our system. However, since our model's critical exponents also differ significantly from the previously studied case of Ref.~\cite{prakash1992structural}, it still falls firmly within its own universality class. This is not so surprising, as our long-range correlations emerge {\it dynamically} due to the presence of memory.

\begin{table}[htbp]
    \centering
    \begin{tabular}{|c|c|c|c|}
    \hline
        Quantity & Our model & \makecell{Uncorrelated \\ site percolation} & \makecell{Long-range correlated \\ percolation, $\lambda=1$} \\\hline
        $p_c$ & $0.5134\pm 0.0005$ & 0.5927 & 0.569 \\\hline
        $\nu_c$ & $1.852\pm 0.001$ & $4/3 = 1.\overline{333}$ & 2.0 \\\hline 
        $\beta_c$ & $0.131 \pm 0.006$ & $5/36 = 0.13\overline{8}$ & 0.21 \\\hline 
        $\gamma_c$ & $3.10 \pm 0.01$ & $43/18 = 2.3\overline{8}$ & 3.6 \\\hline 
        $D$ & $1.924\pm 0.001$ & $91/48 =  1.8958\overline{3}$ & 1.9\\\hline 
    \end{tabular}
    \caption{The percolation threshold $p_c$, critical exponents $\nu_c$, $\beta_c$, and $\gamma_c$, and fractal dimension $D$ at the percolation transition for our model, estimated using finite-size scaling. The reported uncertainties represent one standard deviation, calculated via the bootstrapping method. For reference, the corresponding values for 2D uncorrelated site percolation \cite{christensen2002percolation} and long-range correlated percolation \cite{prakash1992structural} are included. The uncertainties for long-range correlated percolation are not fully reported in \cite{prakash1992structural}, and are thus omitted here. Our system's behavior significantly deviates from both existing models.}
    \label{tab:exponents}
\end{table}

Lastly, we note that this percolation transition is closely related to the avalanches reported in the main text. Recall that an avalanche is defined as a collection of spin flips occurring in rapid succession. In other words, an avalanche represents a spatiotemporally connected cluster of flipping events. By characterizing a flipping event by its lattice site $(x, y)$ {\it and} the time window of length $\Delta_{tw}$ within which it falls, we can further establish a 3D (2 spatial + 1 temporal dimensions) percolation problem, based on the connectivity described in the section on avalanches. This also means that our choice of $\Delta_{tw}$ can be related back to the lattice spacing in the temporal ``direction''.

For each predefined time window $\Delta_{tw}$, the occupation probability on this 3D lattice can be estimated as $p(\Delta_{tw})$, using the function $p(t)$ from Fig.~\ref{fig:p_vs_t} (under the assumption that a spin does not flip more than once within a short time window $\Delta_{tw}$). Thus, when $p(\Delta_{tw})$ approaches the percolation threshold $p_c$ for this custom-defined 3D lattice, we observe a power-law distribution of avalanches (clusters), with the power-law exponent being the Fisher exponent $\tau$ that characterizes the cluster-size distribution.

As demonstrated in the previous sections, our chosen time windows ($\Delta_{tw} = 16$ for $\gamma = 0.01$ and $\Delta_{tw} = 5$ for other cases) resulted in scale invariance in avalanche size distributions. We can now see that these chosen time windows both correspond to a percolation threshold of $p_c = 0.20$, supporting our analogy.

Although $p_c = 0.31$ for 3D {\it cubic} site percolation, our (2+1)-dimensional lattice has distinct connectivity (due to the finite time it takes for information to propagate). However, the reported power-law exponent $\alpha_\gamma \sim 2$ is close to the same exponent for 3D cubic site percolation, $\tau = 2.19$, further validating this analogy. In fact, under the framework of long-range correlated percolation, it has been shown that the Fisher exponent $\tau$ remains unchanged with the introduction of long-range correlations \cite{prakash1992structural}, which is consistent with our numerical results.

\section{Avalanche size distributions in other dimensions}

\begin{figure*}
    \centering
    \includegraphics[width=\textwidth]{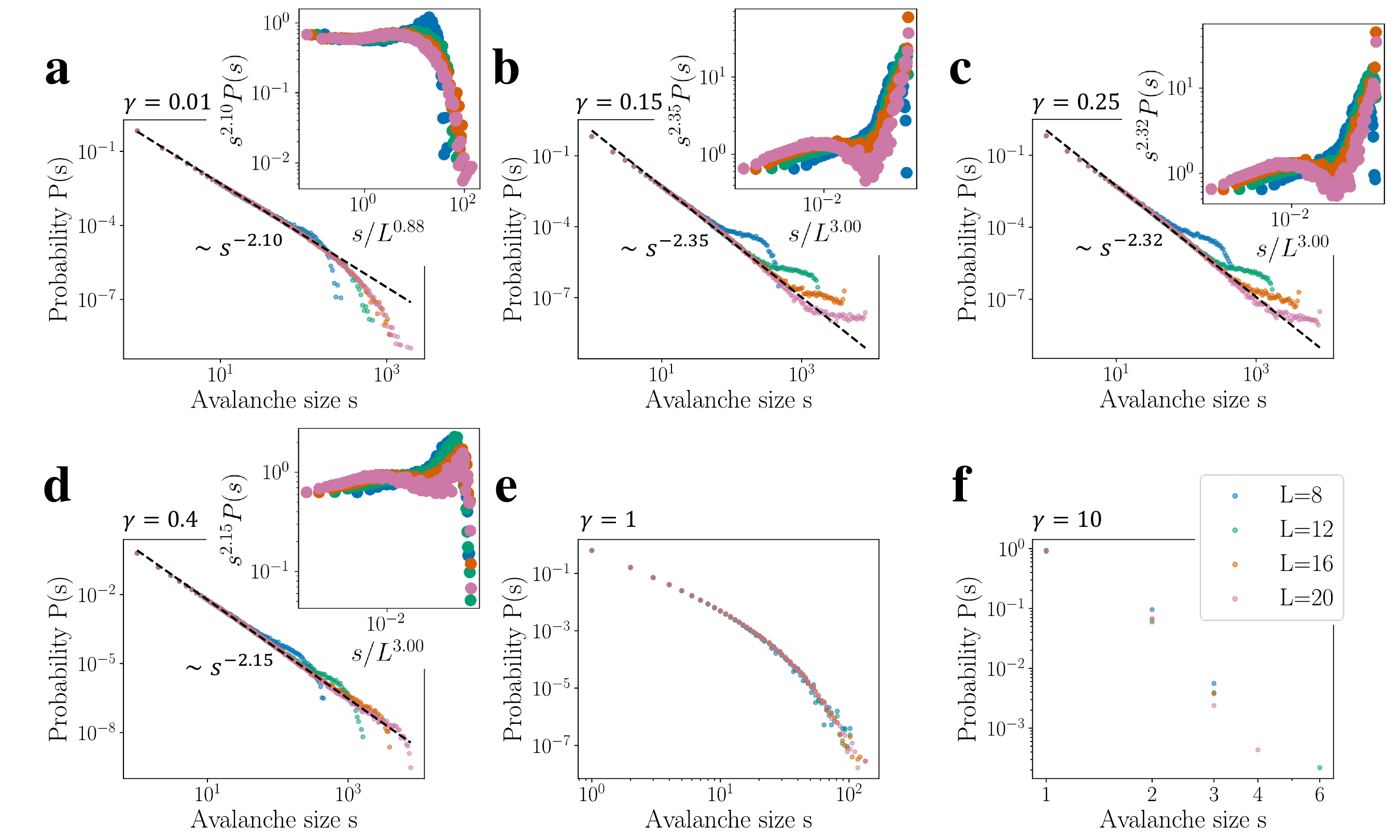}
    \caption{Avalanche size distributions on a 3D lattice for $\gamma \in \{0.01, 0.15, 0.25, 0.4, 1, 10\}$. For each $\gamma$, the distributions are shown for lattice sizes $N \equiv L^3 = 8^3, 12^3, 16^3$, and $20^3$. The insets demonstrate that appropriate rescaling collapses the distributions onto the same curve for different values of $L$, indicating scale invariance. For all cases, the time window $\Delta_{tw}$ is fit by $0.48\gamma^{-1/3}$.
    \label{fig:3d_avalanche}}
\end{figure*}

In the main text, we analyzed a 2D Ising spin glass-inspired model. In this section, we extend the study to 3D. The 1D case is omitted, as the spin glass-inspired model does not apply in 1D due to the absence of frustration. Additionally, the 1D percolation threshold $p_c = 1$ cannot be achieved numerically using our setup.

Our starting point is Eq.~(8) in the main text, applied to a simple cubic lattice. The spin variables $s_i$ reside on the vertices, each with six nearest neighbors, and each neighboring pair of spins, $s_i$ and $s_j$, share a memory variable $x_{ij}$. Following the methodology outlined in the main text, we vary the value of $\gamma$ and analyze the avalanche size distributions for lattice sizes $N \equiv L^3 = 8^3, 12^3, 16^3$, and $20^3$. The results, shown in Fig.~\ref{fig:3d_avalanche}, are consistent with the findings for the 2D case: the system exhibits long-range order when $\gamma \lesssim 0.5$, characterized by a power-law distribution of avalanches. As $\gamma$ increases further, spins become locked in quasiperiodic orbits, leading to the loss of long-range order. We find that taking $\Delta_{tw} = 0.48 \gamma^{-1/3}$ yields reasonable scale-free distributions in this case.

For reference, the power-law exponent for the cluster size distribution in 4D percolation (3 spatial and 1 temporal dimension in our case) is $\tau = 2.31$ \cite{mertens2018percolation}, which agrees with our numerical results shown in Fig.~\ref{fig:3d_avalanche}.

\bibliography{SM_apssamp}